\documentclass[pra,twocolumn,floatfix]{revtex4}

\newcommand{\mrm}{\rm}

\newcommand{\sect}[1]{\paragraph{#1}}

\newcommand{\ns}{{n^{}_S}}

\newcommand{\Is}{I_S}

\newcommand{\btab}{\begin{tabular}}
\newcommand{\etab}{\end{tabular}}

\newcommand{\rhos}{\rho^{}_{\mrm S}}
\newcommand{\rhohs}{\hat{\rho}^{}_{\mrm S}}

\newcommand{\ket}[1]{\mbf{|}#1\mbf{\rangle}}

\newcommand{\rank}{{\bf rank}}

\newcommand{\trace}{{\bf Tr}}

\newcommand{\true}{{\mrm true}}

\newcommand{\real}{{\mrm Re}}
\newcommand{\imag}{{\mrm Im}}
\newcommand{\diag}{{\mrm diag}}

\newcommand{\eps}{\varepsilon}

\newcommand{\alf}{\alpha}

\newcommand{\lam}{\lambda}
\newcommand{\bet}{\beta}

\newcommand{\del}{\delta}
\newcommand{\sig}{\sigma}

\newcommand{\Del}{\Delta}
\newcommand{\Gam}{\Gamma}

\newcommand{\norm}[1]{ \left\| #1 \right\| }

\newcommand{\rmsnorm}[1]{\norm{#1}_{\mrm rms}}

\newcommand{\rhoh}{\hat{\rho}}

\newcommand{\eg}{\emph{e.g.}}
\newcommand{\ie}{\emph{i.e.}}

\newcommand{\bquem}{\begin{quote}\begin{em}}
\newcommand{\equem}{\end{em}\end{quote}}

\newcommand{\blist}{\begin{description}}
\newcommand{\elist}{\end{description}}

\newcommand{\bquote}{\begin{quote}}
\newcommand{\equote}{\end{quote}}

\newcommand{\ben}{\begin{enumerate}}
\newcommand{\een}{\end{enumerate}}

\newcommand{\bit}{\begin{itemize}}
\newcommand{\eit}{\end{itemize}}

\newcommand{\bea}{\begin{array}}
\newcommand{\eea}{\end{array}}

\newcommand{\bds}{\begin{displaystyle}}
\newcommand{\eds}{\end{displaystyle}}

\newcommand{\Rbf}{{\mathbf R}}
\newcommand{\Cbf}{{\mathbf C}}

\newcommand{\ds}{\displaystyle}

\newcommand{\mbf}[1]{\mbox{\boldmath $#1$}}

\newcommand{\refeq}[1]{(\ref{eq:#1})}

\newcommand{\set}[2]{ \left\{ \,#1\, \left| \,#2\, \right.\right\} }

\newcommand{\seqq}[2]{ \Big\{ #1\ \Big\} }

\makeatletter
\def\beq{\@ifnextchar 
[{\@tempswatrue\@beq}{\@tempswafalse\@beq[]}}
\def\@beq[#1]{\begin{equation}\edef\@tmparg{#1}\ifx\@tmparg\@e
mpty \else
	\label{#1}\fi}
\newcommand{\eeq}{\end{equation}}
\makeatother 
\newcommand{\beqaa}{\begin{eqnarray*}}
\newcommand{\eeqaa}{\end{eqnarray*}}

\newcommand{\beqa}{\begin{eqnarray}}
\newcommand{\eeqa}{\end{eqnarray}}

\newcommand{\bc}{\begin{center}}
\newcommand{\ec}{\end{center}}

\usepackage{times}
\usepackage{amssymb}
\usepackage{amsmath}
\usepackage{amsfonts}
\usepackage[dvips]{epsfig}
\usepackage{psfrag}

\newcommand{\normone}[1]{\norm{#1}_{\ellone}}
\newcommand{\pbf}{p_{\rm bf}}
\newcommand{\figpm}{Fig.\ref{fig:x4x4}}

\newcommand{\crao}{Cram\'{e}r-Rao\ }
\newcommand{\rwtalg}{\refeq{rwt1}-\refeq{rwt2}\ }
\newcommand{\elltwo}{{\ell_2}}
\newcommand{\ellone}{{\ell_1}}

\newcommand{\ideal}{{\rm ideal}}

\newcommand{\figest}{Fig.\ref{fig:est4x4}}
\newcommand{\est}{{\rm est}}

\newcommand{\xfeas}{\mbox{$X$ satisfies \refeq{xfeas}}}

\renewcommand{\sect}[1]{\emph{#1}.---}

\newcommand{\Gamb}{\bar{\Gam}}
\newcommand{\gpx}{{\cal G}}
\renewcommand{\Is}{I_n}
\renewcommand{\ns}{{n}}

\newcommand{\nsfour}{{n^4}}

\newcommand{\sumts}{\textstyle\sum}

\renewcommand{\rmsnorm}[1]{\norm{#1}_{\rm rms}}
\newcommand{\emp}{{{\rm emp}}} 
\newcommand{\niouticfg}{N_{\iout\icfg}}
\newcommand{\qs}{{q}}
\newcommand{\nsqpt}{n^4-\nsns}
\renewcommand{\diag}{\mbox{\rm diag}}
\renewcommand{\rhos}{\rho}
\renewcommand{\rhohs}{\rhoh}

\newcommand{\nsns}{{n^2}}

\newcommand{\Cbfnsns}{\Cbf^{\ns\times\ns}}

\renewcommand{\set}[2]{ \left\{#1 \left| #2 \right.\right\} }

\newcommand{\iout}{{i}}
\newcommand{\icfg}{{k}}

\renewcommand{\seqq}[3]{\{#1\}_{#2}^{#3}}

\renewcommand{\rank}{\mbox{\rm rank}}

\newcommand{\ls}{ {\rm LS} }
\newcommand{\ml}{ {\rm ML} }

\newcommand{\nout}{{n_{\rm out}}}
\newcommand{\ncfg}{{n_{\rm cfg}}}
\newcommand{\nex}{N}

\begin{document}

\title{
Quantum~Process~Tomography~via~$\ell_1$-norm Minimization}

\author{Robert L. Kosut}
\affiliation{SC Solutions, Sunnyvale, CA 94085 
{\rm({\tt kosut@scsolutions.com})}}

%%%%%%%%%%%%%%%%%%%%%%%
\begin{abstract}

For an initially well designed but imperfect quantum information
system, the process matrix is almost sparse in an appropriate
basis. Existing theory and associated computational methods
($\ell_1$-norm minimization) for reconstructing sparse signals
establish conditions under which the sparse signal can be perfectly
reconstructed from a very limited number of measurements (resources).
%%%%%%%%%%%%%%%%%%%%%
\iffalse 
In some instances it can be shown that the number of measurements
scales logarithmically with the problem dimension.  For QPT this
\emph{could} mean that resources scale linearly rather than
exponentially with the number of qubits. A direct extension to quantum
process tomography of the $\ell_1$-norm minimization theory has not
yet emerged.
\fi
%%%%%%%%%%%%%%%%%%5
Although a direct extension to quantum process tomography of the
$\ell_1$-norm minimization theory has not yet emerged, the numerical
examples presented here, which apply $\ell_1$-norm minimization to
quantum process tomography, show a significant reduction in resources
to achieve a desired estimation accuracy over existing methods.

\end{abstract}

\maketitle
%%%%%%%%%%%%%%%%%%%%
%\sect{Introduction}
%
\emph{Quantum process tomography} (QPT) refers to the use of measured
data to estimate the dynamics of a quantum system
\cite{NielsenC:00,DarianoP:01}.  Unfortunately, in the general case,
the dimension of the parameter space for QPT can be prohibitive,
scaling exponentially with the number of qubits. This in turn places
the same burden on resources, \eg, the number of applied inputs,
measurement outcomes, and experiments to achieve a desired accuracy,
as well as estimation computational complexity. A number of approaches
have been developed to alleviate this burden. Of note are the various
forms of ancilla assisted QPT (see \cite{MohseniRL:08} for a review),
and the use of symmetrisation to estimate selected process properties
\cite{EmersonETAL:07}. Here we present a method which can be used
either alone or in conjunction with any of the aforementioned
approaches.  The underlying premise is that for an intially well
engineered design, the object that describes the quantum dynamics, the
\emph{process matrix}, will be \emph{almost sparse} in the appropriate
basis.  Certainly in the ideal case of a perfect unitary channel, in
the corresponding ideal basis, the process matrix is maximally sparse,
\ie, it has a \emph{single} non-zero element. Since environmental
interactions cannot be totally eliminated, the actual process matrix
in this ideal basis will be populated with many small elements, and
thus, is almost sparse.

These are the conditions under which methods using $\ellone$-norm
minmization -- often referred to as \emph{Compressive Sensing} -- are
applicable \cite{Donoho:06,CandesRT:06,CandesWB:08}. Specifically, for
a class of incomplete linear measurement equations ($y=Ax,\
A\in\Rbf^{m\times n},\, m\ll n$), constrained $\ell_1$-norm
minimization (minimize $\normone{x}$ subject to $y=Ax$), a convex
optimization problem, can perfectly estimate the sparse variable
$x$. These methods also work very well for systems which do not
satisfy the theoretical conditions, \ie, for almost sparse variables
and with measurement noise.

The underlying theory of $\ellone$ minimization shows that under
certain conditions on the matrix $A$, to realize perfect recovery, the
number of measurements, $m$, scales with the product of the log of the
number of variables $n$ and the sparsity.  Since QPT parameters are
linear in probability outcomes, and scale exponentially with the
number of qubits, this approach heralds a possible linear scaling with
qubits. The theory, however, has not as yet been extended to
QPT. \emph{The numerical examples here are not meant to lend support
to this scaling as they are only presented for the two-qubit case.}
The examples do, however, show more than an order of magnitude savings
in resources over a standard constrained least-squares estimation
using a complete set of measurements, \ie, $\rank(A)\geq n$.

%%%%%%%%%%%%%%%%%%%%%%
The paper is organized as follows: QPT formalism is described next,
followed by a discussion of the genesis of process matrix (almost)
sparsity. A form of the $\ell_1$ minimization for QPT is then
presented followed by numerical examples and some concluding remarks.

%%%%%%%%%%%%%%%%%%%%%%%%%%%%%%
\sect{QPT Formalism}
Recall that the state-to-state dynamics of an \emph{open}
finite-dimensional quantum system can be described in the following
canonical form \cite{NielsenC:00}:
\beq[eq:osrx]
\rhohs
=
\sumts_{\alf,\bet=1}^{\nsns} 
X_{\alf\bet} \Gam_\alf \rhos \Gam_\bet^\dag
\eeq
where $\rhos,\ \rhohs\in\Cbfnsns$ are the input and output state,
respectively, of dimension $\ns$, $X_{\alf\bet}$ are the elements of
the $\nsns\times\nsns$ \emph{process matrix} $X$, and the matrices
$\Gam_\alf$ form an orthonormal basis set for $\ns\times\ns$ complex
matrices:
\beq[eq:gam]
\set{\Gam_\alf\in\Cbfnsns}{\trace\ \Gam_\alf^\dag\Gam_\bet
= \del_{\alf\bet}, \alf,\bet=1,\ldots,\nsns}
\eeq
It is assumed that the quantum system to be estimated is
\emph{completely positive and trace preserving} (CPTP). The set of
feasible process matrices is then restricted to the convex set
\cite{KosutWR:04,BoydV:04},
\beq[eq:xfeas]
\bea{l} 
\mbox{$X\geq 0$ (positive semidefinite)}
\\ 
\sumts_{\alf,\bet=1}^\nsns
X_{\alf\bet}\Gam_\bet^\dag\Gam_\alf=\Is 
\eea
\eeq
It follows from \refeq{xfeas} that the number of real parameters in
the process matrix is $\nsqpt$.  For $\qs$ qubits $\ns=2^\qs$, hence,
scaling with parameters is exponential in the number of qubits.
%\eg, if $\qs=[1,\ 2,\ 3,\ 4]$, then $\nsqpt=[12,\ 240,\ 4032,\ 65280]$.

%%%%%%%%%%%%%%%%%%%%%%%%
\sect{Collecting data}
A common method for collecting data from a quantum system is via
repeated identical experiments.  Denote by $\iout=1,\ldots,\nout$ the
distinct \emph{outcomes}, and by $\icfg=1,\ldots,\ncfg$ the
experimental \emph{configurations}, \eg, any ``knobs'' associated with
state inputs and/or measurement devices.  The measurement outcomes are
recorded from identical experiments in each configuration $\icfg$
repeated $N_\icfg$ times.  Let $\niouticfg$ denote the number of times
out of $N_\icfg$ that outcome $\iout$ occurred in configuration
$\icfg$. The QPT data are the recorded outcome counts,
\beq[eq:nik]
\bea{l}
\set{\niouticfg}{\iout = 1,\ldots,\nout,\ 
\icfg=1,\ldots,\ncfg}
%\\
%\nex = \sumts_{\icfg=1}^\ncfg N_\icfg
%=
%\sumts_{\icfg=1}^\ncfg \sumts_{\iout=1}^\nout \niouticfg
\eea
\eeq
where $\nex=\sumts_{\icfg=1}^\ncfg N_\icfg =\sumts_{\icfg=1}^\ncfg
\sumts_{\iout=1}^\nout \niouticfg$ is the total number of
experiments.

%%%%%%%%%%%%%%%%%%%%%%
\sect{Estimating the process matrix}
An {\em empirical estimate} of the probability of measuring outcome
$\iout$ in configuration $\icfg$ can be obtained from \refeq{nik} as,
\beq[eq:pik emp]
p_{\iout\icfg}^\emp
= 
\niouticfg/N_\icfg
\eeq
From the Born Rule the \emph{model probability} of outcome $i$ given
configuration $k$ with observable $M_{\iout\icfg}$ is,
$%\beq[eq:pik]
p_{\iout\icfg} = \trace\ M_{\iout\icfg}\rhohs_\icfg
$, %\eeq
where from \refeq{osrx}, $\rhohs_k= \sumts_{\alf,\bet=1}^{\nsns}
X_{\alf\bet} \Gam_\alf \rhos_k \Gam_\bet^\dag $. In terms of the
process matrix $X$, the Born rule then becomes, 
\beq[eq:pikx] 
\bea{rcl}
p_{ik}(X) &=& \trace\ G_{\iout\icfg}X
\\
(G_{\iout\icfg})_{\alf\bet}
&=&
\trace\ \Gam_\bet^\dag M_{\iout\icfg}\Gam_\alf\rhos_k 
\eea 
\eeq
The $\nout\ncfg$ matrices $G_{\iout\icfg}\in\Cbfnsns$ capture the
effect of measurements in the matrix basis set \refeq{gam}.  For each
outcome $\iout$, the complete set of configurations is the combination
of all these matrices and the input states:
$\seqq{\rhos_\icfg,G_{\iout\icfg}}{\icfg=1}{\ncfg}$.

A process matrix estimate can be obtained by minimizing the difference
between the empirical probability estimates $p_{ik}^\emp$ and the
model probabilities $p_{ik}(X)$ subject to the feasibility constraint
\refeq{xfeas}. Using a ``least-squares'' measure of probability error
leads to estimating the process matrix by solving the optimization
problem:
\beq[eq:xls]
\bea{ll}
\mbox{minimize} 
& 
V_\ls(X) = \sumts_{\iout ,\icfg} 
\left( p_{ik}^\emp-p_{ik}(X) \right)^2
\\ 
\mbox{subject to} 
& 
\xfeas
\eea
\eeq
Because the outcomes of each experiment are independent, a maximum
likelihood approach can also be considered, \ie,
\beq[eq:xml]
\bea{ll}
\mbox{minimize}
&
V_\ml(X) = -\sumts_{\iout ,\icfg} \niouticfg \log p_{ik}(X) 
\\
\mbox{subject to}
&
\xfeas
\eea
\eeq
Both \refeq{xls} and \refeq{xml} are convex optimization problems with
the optimization variables being the elements of $X$
\cite{KosutWR:04,BoydV:04}. The resulting solution (estimate) will
always be CPTP \refeq{xfeas}. Unfortunately, as already mentioned, the
dimension of the parameter space ($\nsqpt,n=2^q$) can severely strain
resources to the point of impracticality. To see this more clearly,
let the linear relation in \refeq{pikx} between the $\nout\ncfg$ model
probability outcomes and the $\nsfour$ elements of the process matrix
be represented by an $\nout\ncfg\times\nsfour$ matrix $\gpx$, \ie,
\beq[eq:gpx]
\vec{p}=\gpx \vec{X}
\eeq
where $\vec{p},\vec{X}$ are vectors formed from the $p_{ik}$ and
elements of $X$, respectively.  Accounting for the $\ns^2$ linear
constraints in \refeq{xfeas}, $X$ can be recovered from either
\refeq{xls} or \refeq{xml} to within any desired accuracy by using
enough data ($\nex$ in \refeq{nik} sufficiently large), provided that
$\rank(\gpx) \geq \nout\ncfg \geq \nsqpt$. Therefore it would seem
that the resources, $\nout\ncfg$, must also scale exponentially with
the number of qubits. This, however, is not the case when the process
matrix is almost sparse and where the sparsity pattern is not
known\footnote{
A known sparsity pattern can arise from the underlying dynamics,
thereby inherently increasing QPT efficiency \cite{MohseniR:08}.
}.

%%%%%%%%%%%%%%%%%%%%%
\sect{Almost sparsity of the process matrix}
With no noise the ideal channel $\rho\to\rhoh$ for a quantum
information system is a unitary, \ie, $\rhoh=U\rho U^\dag$. Let
$\seqq{\Gamb_\alf\in\Cbfnsns}{\alf=1}{\nsns}$ denote the
``Natural-Basis'' for matrices in $\Cbfnsns$, \ie, each basis matrix
has a single non-zero element of one. In this basis, the process
matrix associated with the ideal unitary channel has the rank-1 form,
$X_\ideal=xx^\dag$ with $x\in\Cbf^{\ns^2},\ x^\dag x=\ns$. A singular
value decomposition (SVD) gives $X_\ideal=V\diag(n,0,\ldots,0)V^\dag$
with $V\in\Cbfnsns$ a unitary.  An equivalent process matrix can be
formed from the SVD in what is referred to here as the
``Ideal/SVD-Basis,''
$
\seqq{
\Gam_\alf=
\sumts_{\alf'=1}^{\nsns} V_{\alf'\alf} \Gamb_{\alf'}
\in\Cbfnsns}{\alf=1}{\nsns}
$, 
The equivalent process matrix, in this basis, denoted by $X_\ideal$, is
maximally sparse with a single non-zero element, specifically,
$(X_\ideal)_{11}=\ns$.  As will always be the case, the actual channel
will be a perturbation of the ideal unitary. If the noise source is
small then the process matrix in the nominal basis will be almost
sparse.

%%%%%%%%%%%%%%%%%%%%%
\sect{Example: Noisy two-qubit memory}
Consider a system which is ideally a two-qubit quantum memory, thus
$U=I_4,\ns=4$.  Suppose the actual system is a perturbation of
identity by independent bit-flip errors in each channel occurring with
probability $\pbf$. For $\pbf=0.05$ and $\pbf=0.2$, the respective
channel fidelities are about 0.90 and 0.64, which for quantum
information processing would need to be discovered by QPT and then
corrected for the device to ever work.  Referring to \figpm, in the
Natural-Basis, \figpm(a), the ideal $16\times 16$ process matrix has
16 non-zero elements out of $256$, all of magnitude one. Using the
Ideal/SVD-Basis the corresponding process matrix as shown in
\figpm(b)\ has a \emph{single} non-zero element of magnitude $\ns=4$
-- it is clearly maximally sparse. \figpm(c)-(d) and (e)-(f),
respectively, show the effect of the two $\pbf$ levels in the two
basis sets. In the Ideal/SVD-basis \figpm(d) and (f) show that the
actual (noisy) process matrices are almost sparse.

%%%%%%%%%%%%%%%%%%%%%%%%%%%%%%
\begin{figure}[t]
\btab{c}
\hspace{-.25in}
\epsfig{file=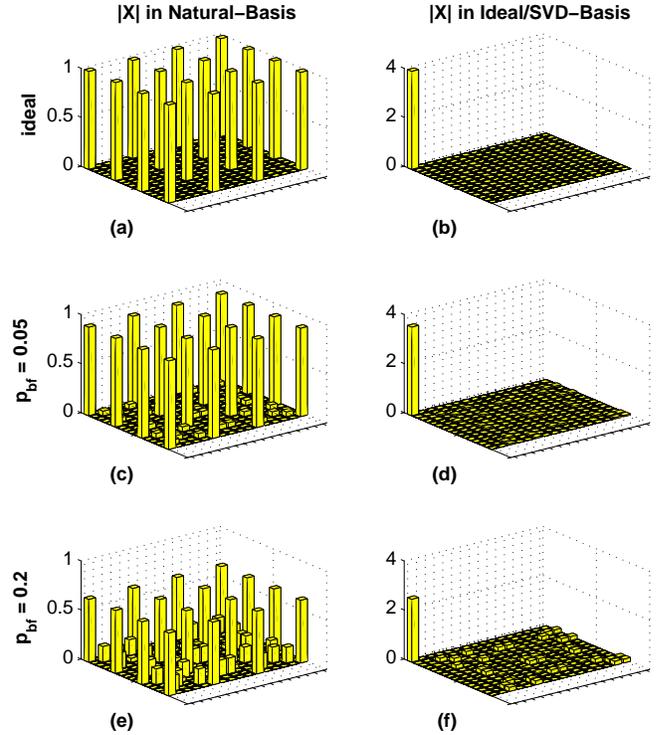,
width=3.5in}%4in}
\etab
%\vspace{-.75in}
\vspace{-3ex}
\caption{
Absolute values of the elements of the process matrix
$X\in\Cbf^{16\times 16}$ for: (a) ideal in the Natural-Basis; (b)
ideal in Ideal/SVD-Basis; (c) actual ($\pbf=0.05$) in
Natural-Basis, (d) actual ($\pbf=0.05$) in Ideal/SVD-Basis;(e) actual
($\pbf=0.2$) in Natural-Basis, (f) actual ($\pbf=0.2$) in
Ideal/SVD-Basis.
}
\label{fig:x4x4}
\end{figure}
%%%%%%%%%%%%%%%%%%%%%%%%%%%%%%%%%

%%%%%%%%%%%%%%%%%%%%%%%%%%%
\sect{Sparsity minimization}
A known heuristic for minimizing sparsity without knowing the sparsity
pattern, and also accruing the benefit of using fewer resources, is to
minimize the $\ell_1$-norm of the vector of variables
\cite{Donoho:06,CandesRT:06,BoydV:04}. For QPT the equivalent $\ell_1$
norm is defined here as the sum of the absolute values of the real and
imaginary parts of each element of the process matrix. There are many
related approaches to incorporate this norm. For example, an estimate
of $X$ can be obtained by solving the following convex optimization
problem:\footnote{
There are many alternatives to \refeq{1norm}, \eg,
$
\mbox{
minimize
$\normone{X}+\lam V(X)$
subject to
$\xfeas$
}
$
or 
$
\mbox{
minimize
$V(X)$
subject to $\normone{X}\leq s$,
$\xfeas$
}
$.
}
\beq[eq:1norm]
\bea{ll}
\mbox{minimize}
&
\normone{X}
\equiv \sum_{\alf,\bet=1}^\nsns 
(|\real\ X_{\alf\bet}| + |\imag\ X_{\alf\bet}|)
\\
\mbox{subject to}
&
V(X) \leq \sig,
\xfeas
\eea
\eeq
with, \eg, $V(X)$ from \refeq{xls} or \refeq{xml}.  The optimization
parameter $\sig$ is used to regulate the tradeoff between fitting $X$
to the data by minimizing $V(X)$ vs. minimizing the sparsity of $X$
via the $\ell_1$-norm. Selecting $\sig$ is often done by averaging
$V(X)$ over a series of surrogates for $X$ obtained from anticipated
scenarios or iterating estimation and experiment design, \eg,
\cite{KosutWR:04}. 
 
In the examples to follow we use the modification of \refeq{1norm}
suggested in \cite{CandesWB:08}, referred to there as
``$\ell_1$-reweighted minimization.''  In this approach a weighted
$\ell_1$-norm is used with the weights determined iteratively. The
algorithm described in \cite{CandesWB:08} is:
\begin{description}
\setlength{\itemsep}{-1ex}
\item {\bf Initialize} $\sig> 0,\ \eps>0,\ W=I_{\nsfour}$
\item {\bf Repeat}
\vspace{-1ex}
	\ben
\setlength{\itemsep}{-1ex}
\item \emph{Solve for $X$}
\beq[eq:rwt1]
\bea{ll}
\mbox{minimize}
&
\normone{WX}
\\
\mbox{subject to}
&
V(X) \leq \sig,
\xfeas
\eea
\eeq
\item \emph{Update weights}
\beq[eq:rwt2]
\bea{rcl}
W &=& 
\ds
{\rm diag}
\left(
1/(|x_1|+\epsilon),\ldots,1/(|x_{\nsfour}|+\epsilon)
\right)
\\
x &=& \vec{X}
\eea
\eeq
	\een
\item {\bf Until} \emph{convergence} --  the objective stops
decreasing or a maximum number of iterations is reached.
\end{description}
In each of the examples to follow the procedure for QPT is: (i) solve
\refeq{xls} to obtain $X_\elltwo$; (ii) set $\sig=1.3\ V(X_\elltwo)$;
(iii) solve the reweighting algorithm \rwtalg for $X_\ellone$.

%%%%%%%%%%%%%%%%%%%%%%%%%%%%%%%%%%%%%
\sect{Example: QPT of noisy two-qubit memory}
For the systems from the example in \figpm, the inputs and
measurements are selected from the set of two-qubit states: $\ket{a}$,
$\ket{+}=(\ket{a}+\ket{b})/\sqrt{2}$,
$\ket{-}=(\ket{a}-i\ket{b})/\sqrt{2}$ with $a,b =
1,\ldots,16$. Specifically, the available set of states are the 16
columns of the matrices,
\beq[eq:ketset]
\begin{tiny}
\left[
\bea{cccc}
1& 0& 0& 0\\
0& 1& 0& 0\\
0& 0& 1& 0\\
0& 0& 0& 1
\eea
\right],\
{\scriptstyle \frac{1}{\sqrt{2}}}
\left[
\bea{cccccc}
1&     1&     1&     0&     0&     0\\
1&     0&     0&     1&     1&     0\\
0&     1&     0&     1&     0&     1\\
0&     0&     1&     0&     1&     1
\eea
\right],\
{\scriptstyle \frac{1}{\sqrt{2}}}
\left[
\bea{cccccc}
1&     1&     1&     0&     0&     0\\
-i&     0&     0&     1&     1&     0\\
0&     -i&     0&     -i&     0&     1\\
0&     0&     -i&     0&     -i&    -i
\eea
\right]
\end{tiny}
\eeq
Considering only coincident input/measurement counts
\cite{ObrienETAL:04}, the relevant probability outcomes \refeq{pikx}
are,
\beq[eq:pabx] 
\bea{rcl}
p_{ab}(X) &=& g_{ab}^\dag X g_{ab},\ X\in\Cbf^{16\times 16}
\\ 
(g_{ab})_\alf
&=& \phi_a^\dag \Gam_\alf \phi_b,\
\alf=1,\ldots,16
\eea 
\eeq
with $\phi_a,\phi_b\, (a,b)\in\{1,\ldots,16\}$ the selected columns
of \refeq{ketset}.

\figest\ shows the error in estimating the process matrix $\Del
X=X_\true-X_\est$ as measured by the RMS matrix norm $\rmsnorm{\Del
X}=(1/n)(\trace\ \Del X^\dag \Del X)^{1/2}$ vs. the number of
experiments per input selected from the set \refeq{ketset} \footnote{
The number of experiments per input/measurement configuration here is
chosen uniformly.  An optimal (non-uniform) choice which minimizes the
\crao lower bound can be cast as a convex optimization problem
\cite{KosutWR:04}.
}.
The results shown are from simulations described in the caption. 
%% The error bars arise from 50 runs for each of the QPT estimates.
%% As this error measure depends on the unknown true 
%% system, it is used here to post-evaluate performance.
%%%%%%%%%%%%%%%%%%%%%%%%%%%%%%%%%%%%%%%
\psfrag{inf}{$\infty$}
\psfrag{L2 256x256}{$X_{\ell_2}$, $\gpx\in\Cbf^{256\times 256}$}
\psfrag{L1 36x256}{$X_{\ell_1}$, $\gpx\in\Cbf^{36\times 256}$}
\begin{figure}[t]
\btab{c}
%\\
%\hspace{.5in}
\epsfig{file=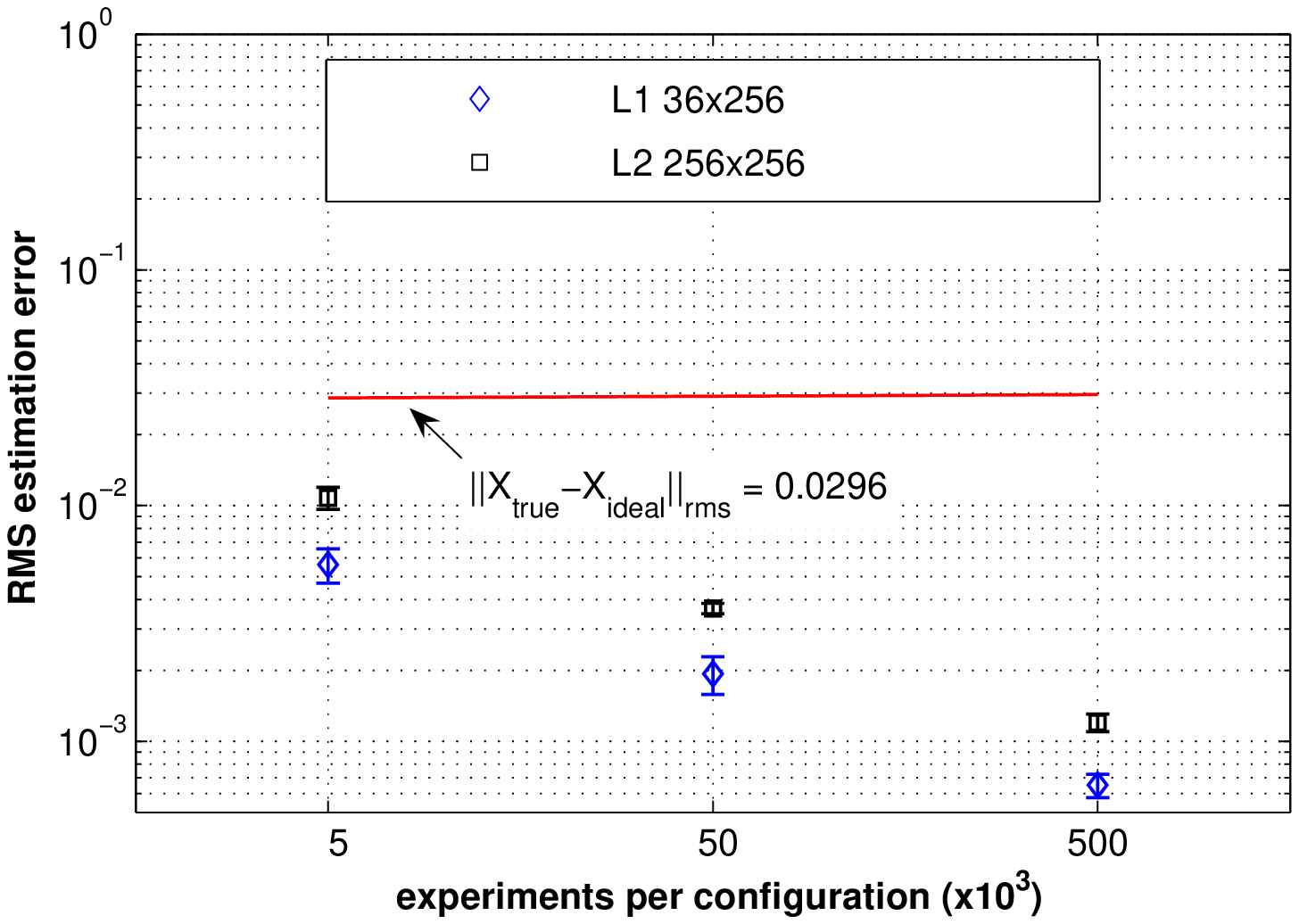,
%width=3in}
height=2in}
\\
(a) $p_{\rm bf}=0.05$
\\
\epsfig{file=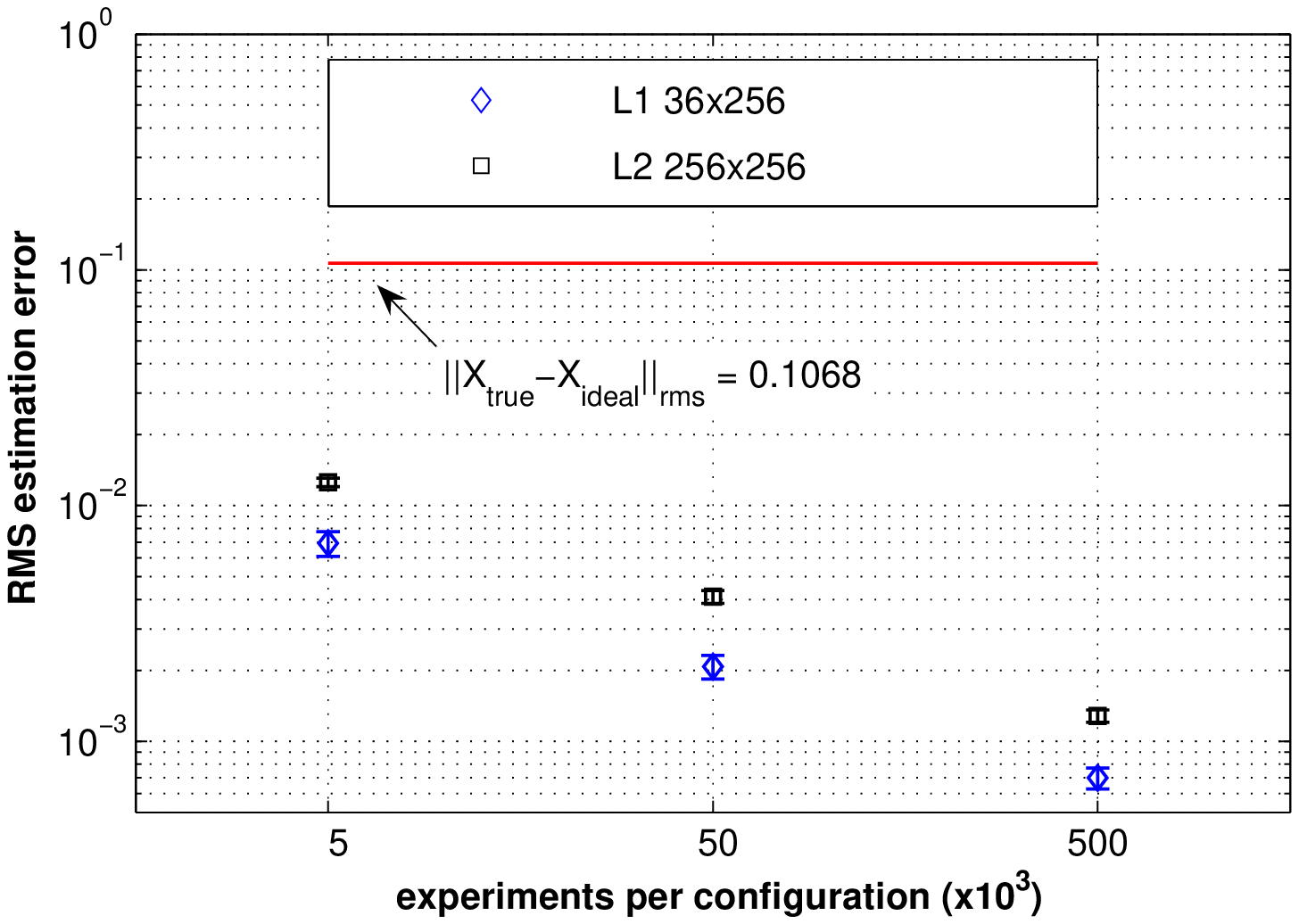,
%width=3in}
height=2in}
\\
(b) $p_{\rm bf}=0.20$
\etab
%\vspace{-.5in}
\caption{
RMS estimation error $\rmsnorm{X_\true-X_\est}$ vs. number of
experiments per configuration: selected columns of
\refeq{ketset}. Error bars show the deviation from 50 runs at each
setting.
\newline
\textbf{$\ell_2$-minimization ($\Box$):} $X_\est=X_\elltwo$ is from
\refeq{xls} using all 16 input/output combinations. This gives a
matrix $\gpx\in\Cbf^{256\times 256}$ as defined in \refeq{gpx} which
is full rank, \ie, $\rank(\gpx)=256$.
\newline
\textbf{$\ell_1$-minimization ($\Diamond$):} $X_\est=X_\ellone$ is from
\rwtalg using 6 inputs and 6 measurements obtained from the columns of
the second matrix in \refeq{ketset}.  This gives
$\gpx\in\Cbf^{36\times 256}$ which is full rank, \ie,
$\rank(\gpx)=36$.
}
\label{fig:est4x4}
\end{figure}
%%%%%%%%%%%%%%%%%%%%%%%%%%%%%%%%%%%%%%%%
%

The benefit of $\ell_1$-minimization compared to the standard
$\ell_2$-minimization is seen most clearly with small amounts of data
from highly incomplete measurements.  For example, for $\pbf=0.05$
[\figest(a)], at $50\times 10^3$ experiments per input for the
6-input/6-output configuration $(\gpx\in\Cbf^{36\times 256}$) the
$\ell_1$ RMS estimation error is $0.0019$. Compare this to the
$\ell_2$ error of $0.0012$ at $500\times 10^3$ experiments per input
for the 16-input/16-output configuration ($\gpx\in\Cbf^{256\times
256}$). The latter improvement can be attributed mostly to the 10-fold
increase in the number of experiments per input. The additional
resources to achieve this are significant, \ie, 16 inputs for $\ell_2$
vs. 6 for $\ell_1$, and additionally, an increase in the \emph{total}
number of experiments from $6\times 50\times 10^3$ to $16\times
500\times 10^3$.  It is certainly not intuitive that to estimate the
240 parameters of the process matrix, the clearly incomplete set of
measurements using only 36 outcomes ($\Diamond$ in \figest) could
produce results not only similar to, but for each number of
experiments per input, even better than the full input case with all
256 combinations of inputs and measurements ($\Box$ in \figest). As
seen the $\ell_1$ error is about 1/2 the $\ell_2$ error. Also,
reweighting reduced the (unweighted) $\ell_1$ error by 1/2-1/3.

Comparing the estimation errors with the error between the actual and
ideal (solid lines in \figest) suggests that at least $50\times 10^3$
experiments per input are needed to achieve a sufficient post-QPT
error correction towards the ideal unitary.  \figest\ also reveals
that the estimation errors are very similar for both levels of
bit-flip error, $\pbf\in\{0.05,0.20\}$. This is explained by the \crao
bound which defines the asymptotic error of any unbiased estimator,
\ie, the RMS decays as $\Del/\sqrt{N}$. Here $\Del$ is effectively the
error between the empirical \refeq{pik emp} and actual \refeq{pikx}
probabilities which by definition is of order one; this provides a
reasonable fit to the data in \figest.

%%%%%%%%%%%%%%%%%%
\sect{Infinite data} With infinite data the measurements are
effectively noise-free, so the empirical probability estimates are
equivalent to the true probabilities.  Infinite data estimates are
obtained by solving \refeq{xls} and \rwtalg with the constraint
$V(X)\leq\sig$ replaced by the linear equality constraint
$p_{ik}(X)=p_{ik}(X_\true)$. 
%%
\iffalse
From \refeq{pikx}, this is equivalent to $\trace\
G_{ik}(X-X_\true)=0$.
\fi
%%
For the numerical examples here, \refeq{pabx} gives the linear
equality $g_{ab}^\dag(X-X_\true)g_{ab}=0$.

In the examples, both $X_\ellone$ from \rwtalg and $X_\elltwo$ from
\refeq{xls} were numerically equal to $X_\true$.  This is to be
expected for $X_\elltwo$ because of the complete set of 256 full rank
measurements. Almost sparsity makes perfect estimation possible with
the highly incomplete set of 36 measurements.

The infinite data case is useful for evaluating different
configuration strategies in simulation, \ie, consider only those that
result in a good estimate.  

To stress the efficacy of $\ell_1$-minimization as a heuristic for
sparsity, consider replacing the $\ell_1$ norm in \rwtalg with the RMS
norm $\rmsnorm{X}$, which is effectively the $\ell_2$ norm of
$\vec{X}$. Solving the 6-input/6-output case ($\Diamond$ in \figest)
for $\pbf=0.05$ with infinite data gives an RMS error of $0.11$, which
is considerably larger than the error between the actual and ideal of
0.03 (solid line in \figest(a)). The estimate gets even worse with
finite data. This again emphasizes the advantage of $\ell_1$
minimization for sparse signal reconstruction
\cite{Donoho:06,CandesRT:06}.

%%%%%%%%%%%%%%%%%%%%%%%%%%
\sect{Conclusions}
The use of the $\ell_1$-norm minimization methods of Compressive
Sensing \cite{Donoho:06,CandesRT:06,CandesWB:08} appear to apply
equally well to sparse QPT. The examples of sparse process matrices
presented here are meant to represent typical initial imperfect,
designs.  The numerical results illustrate how estimation resource
tradeoffs can be obtained. Additionally, the findings suggest that QPT
resources need not scale exponentially with qubits. In the ideal
case, the theoretical question of showing linear scaling with sparsity
using $\ell_1$ minimization for QPT remains open.

Because $\ell_1$ minimization uses considerably fewer resources than
standard QPT, use in an on-line setting combined with optimal quantum
error correction tuned to the specific QPT errors is compelling, \eg,
\cite{ReimpellW:05,FletcherSW:06,KosutSL:08}. 
%%
\iffalse
The final decision as to whether a particular QPT experiment design
suffices for a choice of resources depends on the intended purpose of
the estimation, and what the necessary level of error is in order to
meet performance.
\fi
%%
Another future direction is in conjunction with Hamiltonian parameter
estimation. Here a bank of estimators can be applied to the data where
each estimator is tuned via the Ideal/SVD-Basis to one of a number of
finite samples of the unknown parameters. Such an approach may prove
useful for a small number of parameters. In quantum metrology often a
single uncertain parameter is to be estimated in an unknown noisy
environment, \eg, \cite{glm:06,DornerETAL:08}.

\iffalse
In QPT, if $X$ has sparsity $s$, and if $\gpx$ is in the known matrix
class, then the rank condition above is replaced with $\rank(\gpx)\geq
m \geq s{\cal O}(\log(\nsqpt))\approx s{\cal O}(q)$. At present this
remains an open question for QPT.
\fi

%\iffalse
\sect{Acknowledgments}
Thanks to A. Gilchrist, I. Walmsley, D. Lidar, H. Rabitz, and
M. Mohseni for suggestions and comments.  The idea of applying
$\ell_1$ minimization to QPT arose during discussions
%% in R.L.K. at the PI Workshop
at \footnote{
\emph{Workshop on Quantum Estimation: Theory and Practice}, 
Aug. 25-30, 2008, Perimeter Institute, Waterloo, Canada.
}
. 
%\fi

%%%%%%%%%%%%%%%%%%%%%%%%%%%%%%%%
%\begin{small}
%\bibliographystyle{plain}
%\bibliographystyle{unsrt}
\bibliographystyle{prsty}
\bibliography{D:/robert/tex/rlk}
%\end{small}
%%%%%%%%%%%%%%%%%%%%%%%%%%%
\end{document}